# OGY Control of Haken Like Systems on Different Poincare Sections

Mozhgan Mombeini
*Institute for Research in Fundamental Sciences (IPM), Tehran, Iran*
*Electrical Enginieering-Control and System*
E-mail: Mzh.mombeini@ipm.ir

**Abstract**
The Chua system, the Lorenz system, the Chen system and The L¨u system are chaotic systems that their state space equations is very similar to Haken system which is a nonlinear model of a optical slow-fast system. These Haken-Like Systems have very similar properties. All have two slow but unstable eigenvalues and one fastest but stable eigenvalue. This lets that an approximation of slow manifold be equivalent with unstable manifold of the system. In other hand, control of discreet model of the system on a defined manifold (Poincare map) is main essence of some important control methods of chaotic systems for example OGY method. Here, by using different methods of defining slow manifold of the H-L systems the efficiency of the OGY control for stabilizing problem investigated.

## 1. Introduction

The OGY (Ott, Grebogi and Yorke) method of control is one of the most interested methods of chaos control [3]. In this paper efficiency of the OGY control on different Poincare sections by defining different manifold for Haken-Like systems is investigated. In part 2 OGY method is introduced. In part 3 a discussion on a proper manifold for Poincare map is done. Part 4 introduces category of Haken Like systems. Part 5 includes motivation of this work and the methodology of using slow manifold in the sense of singular perturbation as proper manifold. Part 6 shows simulation results and part 7 is conclusion.

## 2. OGY Method

Let the controlled system be described by the state space equations

$$\dot{x} = f(x,u) \ , x \in R^n , u \in R^m \quad (1)$$

Obtain the desired (goal) trajectory $x_*(t)$ which is a solution with (1) for $u(t) \equiv 0$. Consider a surface (Poincare section) $S = \{x : s(x) = 0\}$ through the given point $x(0) = x_*(0)$, transversally to the trajectory $x_*(t)$. Solving system equation, and intersections of the system with this manifold we can find $x_{k+1} = p(x_k)$, which called the Poincare map. OGY method says that control of this map is equal to control of the full system and that's sufficient to insert control just in vicinity of desired solution that is defined with $\Delta > 0$ sufficiently small. For example, using the linear state feedback $u_k = Cx_k$ by linearization we obtain $\tilde{x}_{k+1} = A\tilde{x}_k + Bu_k$. Then OGY Control Rule is

$$u_k = \begin{cases} C\tilde{x}_k & if (\|\tilde{x}_k\| \leq \Delta \\ 0, & otherwise \end{cases}$$

### 3. Discussion on Proper manifold for Poincare Map

In this part some properties for a proper manifold to contrast discrete model of the system is discussed intuitively.

**3.1 Property1**: It should contain unstable modes (positive real part eigenvalues of linearized model) of the system to be controlled with control rule.

**3.2 Property2:** It should be a good description of the full system and the system trajectories fall in it as sooner as possible to system be controlled.(It should cover and meet almost spaces of the attractor)

**3.3 Property3:** When the control started the controlled Poincare map should have such behavior that keep system in neighborhood of the desired trajectory (we know that chaotic systems are very sensitive to perturbations).

### 4 Haken Like System

Lorenz sysem,

$$\dot{x} = \sigma(y-x), \dot{y} = -xz + rx - y, \dot{z} = xy - bz, \sigma = 10, b = \frac{8}{3}, r = 28$$

Chen system,

$$\dot{x} = \sigma(y-x), \dot{y} = (r-\sigma)x - xz + ry, \dot{z} = xy - bz, \sigma = 35, b = 3, r = 28$$

And L¨u system

$$\dot{x} = \sigma(y-x), \dot{y} = -xz + ry, \dot{z} = xy - bz, \sigma = 36, b = 3, r = 20$$

Are very famous chaotic systems represented in state space equations that are similar to equations of the Haken system:

$$\dot{x} = \sigma(y-x), \dot{y} = xz - y, \dot{z} = b(a - z - xy),$$

Which is nonlinear model of a optical slow-fast system. Here these systems are introduced as Haken Like (H-L) systems. Chua system is another member of this category. Because of that its fixed point is not located in the slow manifold in the sense of singular perturbation excluded from this paper.

**4.1 Singular Perturbation Method for Haken Like systems**

Singularly perturbed systems are class of the systems in state space with

$$\varepsilon\dot{x} = f(x,y), y = g(x,y), x \in R, y \in R^{n-1} \quad (2)$$

where $\varepsilon$ is a small parameter. Systems with these equations are called multi time scale systems. $x$ is fast and $y$ are slow stats. The slow manifold of (2) is defined with $\dot{y} = g(h(y), y),$ where $h(y)$ is solution of $0 = f(x,y),$.

**4.2 Singular Perturbation Method for Haken Like systems**

All of Haken Like systems have approximately a large coefficient $\sigma$ that with a parameter changing $\varepsilon = \frac{1}{\sigma}$ we have a singular perturbation form $\dot{x} = \sigma(y-x), \dot{y} = g(x,y), x \in R, y \in R^2$. Then slow manifold can be produced with $S = x : x - y = 0$. This manifold is zero order approximation of the slow manifold. Using Fenichel theorem and numerical analysis methods first order approximation of this manifold for this class of systems is calculated in others works [1]. In summery first order approximation of slow manifold for Haken Like systems are:

4.2.1 For Lorenz system: $x = y + \varepsilon(yz - 27y)$

4.2.2 For Chen system: $x = y + \varepsilon(yz - 21y)$

4.2.3 For Lu system: $x = y + \varepsilon(yz - 20y)$

**5 OGY Control on the Slow Manifold in the Sense of Singular Perturbation**

In this part tried to introduce slow manifold of the Haken Like systems in the sense of singular perturbation as the proper manifold for the OGY control.

**5.1 Motivation**

In our previous work, using OGY control for chaotic singularly perturbed systems on the slow manifold we found that after inserting each control pulse, system remain for a long time in the desired point without any insertion of a new control pulse[2]. This time can be estimated with $T = \dfrac{1}{\varepsilon}$. This observation is motivation of this paper to extend the idea of control on slow manifold in the sense of singular perturbation to the nonsingular perturbation systems of the Haken Like class.

**5.2 Properties of the Slow Manifold as a Proper Slow Manifold**

Here properties of the slow manifold in the sense of singular perturbation are discussed.

**Property1**: In this class of systems, the fast state ($x$) is the stable and then slow manifold contain unstable modes. Then discrete model of the system contains all unstable modes of the system and satisfying in property1.

**Property2:** System Trajectories come to slow manifold and having a slow dynamic stay there, but because of the nature of chaotic systems they leave the manifold then by stable fast modes come back to the slow manifold rapidly. So slow manifold satisfying in property2.

**Property3**: Slow manifold contain all slow modes of the system. So when entering in its neighborhood, insertion of a control pulse will not through the trajectory out of the control region because of this slow dynamics. So property 3 is satisfied. Then it seems that slow manifold has properties to be a proper Manifold for OGY control.

**5.3 Other ways of Defining of Slow Manifold for Systems**

There are other ways to capture slow manifold for the system. This method does not satisfying all properties of the proper manifold but here used for comparison with proper slow manifold in the sense of singular perturbation.

**Method1:** "On the attractive parts of the phase space (i.e. where jacobian matrix of the system $J(x)$ have a fast eigenvalue), the slow manifold is locally defined by a plane orthogonal to the tangent system's left fast eigenvector"[1,4].

**Method2**: "On the attractive parts of the phase space, let $z_2(x, y, x)$ and $z_3(x, y, x)$ are the two slow eigenvectors associated with the two slow eigenval-ues, $\lambda_2(x)$, and $\lambda_2(x)$ of $J(x)$. The local slow manifold in the neighborhood of $x$ is generated by these two vectors" [1,4].

**6. Numerical Simulation**

Problem of stabilization on fixed point is $(x_{eq}, y_{eq}, z_{eq})$ considered. Error function is defined as

$$error(T) = \frac{1}{T}\int_0^T ((x(t)-x_{eq})^2 + (y(t)-y_{eq})^2 + (z(t)-z_{eq})^2)dt$$

Control effort defined as $error(T) = \frac{1}{T}\int_0^T u(t)^2 dt$, Where $T$ is simulation time. Each simulation repeated 500 turns with different random initial conditions that are saved for other tests. To enrich comparison, different defined manifolds are: manifold with $x$ constant, manifold with $y$ constant, manifold with ortogonl plane to fast modes(sm1), manifold with plane of two slow vector(sm2), singular perturbation manifold(Sms). Table 1 show the result of this simulation. For all three systems the minimum error is for the slow manifold in the sense of singular perturbation but that's not result of maximum control effort.

Table1

| System | Loren | | Chen | | Lu | |
|---|---|---|---|---|---|---|
| Manifold | Effort | Error | Effort | Error | Effort | Error |
| x | 85.6170 | 0.0086 | 128.6943 | 0.0102 | 103.92339 | 0.0082 |
| y | 110.3247 | 0.0139 | 1255 | 0.1478 | 910.7839 | 0.1130 |
| Sm1 | 70.2334 | 0.0163 | 73.7670 | 0.0103 | 115.3081 | 0.0064 |
| Sm2 | 25.0047 | 0.1863 | 96.1802 | 0.0112 | 128.63 | 0.0195 |
| sms | 78.7876 | 0.0072 | 158.2424 | 0.0058 | 161.09 | 0.005 |

**7. Conclusion**

It seems that slow manifold in the sense of singular perturbation results in minimum error for OGY Control, not as a Result of inserting maximum effort, But as a result of selection of the proper manifold to define discrete model of the system.

**Refrences**